\documentclass{SciPost}

\hypersetup{
    colorlinks,
    linkcolor={red!50!black},
    citecolor={blue!50!black},
    urlcolor={blue!80!black}
}

\newcommand{\Ecm}{E_{\text{cm}}}
\newcommand{\gev}{\text{GeV}}
\newcommand{\tev}{\text{TeV}}

\usepackage{easyReview}
\usepackage{cleveref}

\usepackage[compat=1.0.0]{tikz-feynman}
\usepackage[bitstream-charter]{mathdesign}
\usepackage[normalem]{ulem}
\usepackage{multirow}
\usepackage{graphicx}
\usepackage{subcaption}
\usepackage{caption}
\usepackage{booktabs} 

\newcommand{\ch}[1]{{ {{\color{black}{#1}  }}}}

\DeclareSymbolFont{usualmathcal}{OMS}{cmsy}{m}{n}
\DeclareSymbolFontAlphabet{\mathcal}{usualmathcal}

\fancypagestyle{SPstyle}{
\fancyhf{}
\lhead{\colorbox{scipostblue}{\bf \color{white} ~SciPost Physics }}
\rhead{{\bf \color{scipostdeepblue} ~Submission }}

\fancyfoot[C]{\textbf{\thepage}}
}

\begin{document}

\pagestyle{SPstyle}

\begin{center}{\Large \textbf{\color{scipostdeepblue}{
Top quark FCNC  in Randall-Sundrum models: post-LHC allowed rates and searches 
at $e^+e^-$ and $\mu^+ \mu^-$ colliders\\
}}}\end{center}

\begin{center}\textbf{
Sagar Airen\textsuperscript{1$\star$} and
Roberto Franceschini\textsuperscript{2,3$\dagger$} 
}\end{center}

\begin{center}
{\bf 1} Maryland Center for Fundamental Physics, Department of Physics, University of Maryland,
College Park, MD 20742, USA
\\
{\bf 2} Universit\`a degli Studi Roma Tre and INFN, Via della Vasca Navale 84, I-00146, Rome, Italy
\\
{\bf 3} 
Theoretical Physics Department, CERN, 1211 Geneva 23, Switzerland
\\[\baselineskip]
$\star$ \href{mailto:email1}{\small sairen@umd.edu}\,,\quad
$\dagger$ \href{mailto:email2}{\small roberto.franceschini@uniroma3.it}
\end{center}
{\tiny{\hfill CERN-TH-2025-236}}

\section*{\color{scipostdeepblue}{Abstract}}
\textbf{\boldmath{%
We present the sensitivity to  Flavor Changing Neutral Currents (FCNC) in  interactions involving the top quark at future $e^+e^-$ and $\mu^+\mu^-$ machines. We consider the $Ztc$ vertex as well as four-fermion contact interactions involving top and charm quarks.  
To incorporate limits from (HL-)LHC we consider FCNC from  Randall-Sundrum models and we recast LHC searches for the resonances that at the microscopic level give rise to the FCNC effects. We determine the maximal strength of the effective FCNC couplings $Ztc$ coupling allowed by LHC. We find that the LHC currently improves on the limit set by previous machines, e.g. LEP indirect sensitivity to heavy vectors. Future improvements of direct searches at HL-LHC may reach a level equivalent to $BR(t\to c Z)\simeq 10^{-6}$.  We explore the possibility to probe even smaller FCNC coupling strength using an $e^+e^-$ machine at center-of-mass energy suitable for a Higgs factory $E_{cm}\in$~[200,240]~GeV or to probe contact interactions involving top and charm flavors at a high-energy muon collider at $E_{cm}=10$~TeV.
}}

\vspace{\baselineskip}

\noindent\textcolor{white!90!black}{%
\fbox{\parbox{0.975\linewidth}{%
\textcolor{white!40!black}{\begin{tabular}{lr}%
  \begin{minipage}{0.6\textwidth}%
    {\small Copyright attribution to authors. \newline
    This work is a submission to SciPost Physics. \newline
    License information to appear upon publication. \newline
    Publication information to appear upon publication.}
  \end{minipage} & \begin{minipage}{0.4\textwidth}
    {\small Received Date \newline Accepted Date \newline Published Date}%
  \end{minipage}
\end{tabular}}
}}
}


\vspace{10pt}
\noindent\rule{\textwidth}{1pt}
\tableofcontents
\noindent\rule{\textwidth}{1pt}
\vspace{10pt}


\section{Introduction}
\label{sec:intro}
The study of top quark interactions is a major goal for the future generation of lepton colliders. This has been recognized for a long time~\cite{Han:1998yr,Atwood:1997aa,Atwood:1996aa}. As the Large Hadron Collider (LHC) comes to its maturity and $e^{+}e^{-}$ and $\mu^+ \mu^-$ technology improves, for the first time we can envision to directly study top quarks produced in a clean leptonic environment at both circular\cite{FCC:2018evy} and linear colliders~\cite{Abramowicz:2018aa} in the next decades.

In this work we explore the reach of the reaction 
\begin{eqnarray}
\ell^{+}\ell^{-}\to t+j\, 
\label{eq:single-top}    
\end{eqnarray}
that has also been studied in Refs.~\cite{Durieux:2014th,Shi:2019aa
,Khanpour:2014xla}, with particular focus on the case $j=c$. We focus on  energies suitable for a Higgs factory, which is a crucial phase of the  Higgs, top quark, electroweak (HTE) factory program~\cite{Altmann:2025feg}. The Higgs factory stage is normally considered at 240~GeV~\cite{FCC:2018byv,Ai:2024nmn,Monteil:2021ith} or 250~GeV~\cite{ILC:2013jhg}, but we also consider  smaller center-of-mass energies down to 200~GeV, which is just about the reaction energy threshold.

A detailed study of the reach at energies below 240~GeV has been missing for a long time  in the exploration of the top quark physics (see e.g. Ref.~\cite{Han:1998yr} for an early work). In fact, most of the attention has been recently devoted to the possibility to produce top quark pairs at energies close to the threshold of 
$$ e^+e^- \to t \bar{t}\,.$$
Typical energy points for the study of this process range between 350 and 380 or 500 GeV, depending on the machine under consideration\cite{Abramowicz:2018aa,LinearColliderVision:2025hlt,FCC:2018evy}. 
Attaining such energies allows to study the rise of the $t\bar{t}$ cross-section at the threshold, which can give a most precise determination of the top quark mass. Once that energy can be reached, also the $Ztt$ coupling can be studied, possibly highlighting new physics in the top quark sector.

The possibility to reach the threshold for top quark pair production is very exciting, but also very challenging. As a matter of fact, the entire business of making new high energy machines comes with its uncertainties, so that an assessment of less demanding collider options has been carried out~\cite{Anastopoulos:2025jyh} to establish, even if approximately, the physics potential of less luminous or even lower-energy options.
For this reason we believe is timely and important to assess the physics potential for top quark of machines running at energies at or below the usual Higgs factory center-of-mass energy. 

 The rationale for expecting some potential for energies below 240~GeV is that the luminosity that can be attained in circular machines drops roughly as $\mathcal{L}\sim \Ecm^{-4}$ as the center-of-mass energy $\Ecm$ grows. Thus, lowering the center-of-mass energy one can expect to be able to collect more luminosity, hence have more recorded collisions if the signal cross-section does not change too fast, and in particular if it does not drop too fast towards the closure of the phase-space at $\Ecm \simeq m_{t}+m_c\simeq 175\,\gev$.

As clear from our attempt to increase the luminosity by running at lower energy, the exploration of Flavor Changing Neutral Currents (FCNC) at $e^+e^-$ machines depends critically on getting sufficient luminosity to probe the very weak coupling strengths associated to flavor violations. However, other entirely distinct routes exist to probe FCNC of the top quark. For instance one can try to find the microscopic degrees of freedom that generate the flavor violation, or try to highlight flavor violation in contact interactions that  give rise to the same FCNC effects observable at $\ell^+\ell^-$ in top quark production. 

We consider these additional probes of top quark flavor violation and start our study by updating bounds on the resonances that give rise to top quark FCNC in new physics models. Starting from the survey of Snowmass~2013~\cite{TopQuarkWorkingGroup:2013hxj} we identify Randall-Sundrum models as a prime candidate for $Ztc$ couplings. Indeed all the other models considered in that survey had more stringent bounds and are expected to give rise to smaller FCNC. In \Cref{sec:SMEFT} we describe the origin of the $Ztc$ coupling from effective contact interactions in the Warsaw basis~\cite{Grzadkowski:2010es} and in the RS model. 
In \Cref{sec:recast} we provide recast bounds for RS resonances from bounds on the Heavy Vector Triplet (HVT)~\cite{Pappadopulo:2014qza} model at the LHC, estimating as well the sensitivity of HL-LHC, which will set the baseline for future $e^+e^-$ and $\mu^+\mu^-$ machines.

Motivated by the strong bounds in the multi-TeV range from direct searches of the new physics that originates the FCNC top quark interactions, in \Cref{sec:singletop} we assess the potential to probe top quark flavor violation at the HTE factory\cite{Altmann:2025feg}, or a small variation of the current project, and at future multi-TeV muon collider\cite{InternationalMuonCollider:2025sys}, which can be sensitive to even higher effective mass scales and correspondingly to even smaller BRs. 
In \Cref{sec:conclusions} we present a discussion of the results and our conclusions.

\section{SMEFT interactions \label{sec:SMEFT}}
At dimension six, the following Warsaw basis operators~\cite{Grzadkowski:2010es} contribute to the $Ztc$ coupling in the SMEFT,
\begin{eqnarray}\label{eq:smeft_2fcn_op}
\label{eq:higgs-current-fermion-current}
    O_{\phi q}^{1(ij)} = (\phi i \overset{\text{$\leftrightarrow$}}{D_\mu} \phi)(\bar{q}_i\gamma^\mu q_j),
\\
  \label{eq:higgs-current-fermion-current-3}  O_{\phi q}^{3(ij)} = (\phi i \overset{\text{$\leftrightarrow$}}{D^I_\mu} \phi)(\bar{q_i}\gamma^\mu \tau^I q_j),
\\
\label{eq:higgs-current-fermion-current-R}    O_{\phi u}^{1(ij)} = (\phi i \overset{\text{$\leftrightarrow$}}{D_\mu} \phi)(\bar{u_i}\gamma^\mu u_j),
\\ \label{eq:dipole}
    O_{uW}^{(ij)} = \left( \bar{q}_i \sigma^{\mu\nu} \tau^I u_j \right) \tilde{\varphi} W_{\mu\nu}^I, 
\\ \label{eq:dipole-B}
    O_{uB}^{(ij)} = \left( \bar{q}_i \sigma^{\mu\nu} u_j \right) \tilde{\varphi} B_{\mu\nu},
\end{eqnarray}
where $(ij)=(32)$ and $(23)$ give rise to the $Ztc$ coupling of our interest to mediate the scattering eq.~\eqref{eq:single-top}.

These operators can lead to single-top production at $\ell^+\ell^-$ colliders. Following~\cite{Aguilar-Saavedra:2018ksv}, we work with the following linear combination of the coefficient of the operators in the above equation defined as degrees of freedom, which contribute independently to single-top production through $Z$~\cite{Aguilar-Saavedra:2018ksv},

\begin{align}
    \mathcal{C}_{\varphi q}^{-(3+a)} &\equiv  \left\{ C_{\varphi q}^{1(3a)} - C_{\varphi q}^{3(3a)} \right\},  \label{eq:triplet-current-current} \\
    \mathcal{C}_{\varphi u}^{(3+a)} &\equiv   \left\{ C_{\varphi u}^{1(3a)} \right\}, \label{eq:singlet-current-current} \\
    \mathcal{C}_{uZ}^{(3a)} &\equiv \left\{ -s_W C_{uB}^{(3a)} + c_W C_{uW}^{(3a)} \right\}, \label{eq:dipole-Z-LR} \\
    \mathcal{C}_{uZ}^{(a3)} &\equiv \left\{ -s_W C_{uB}^{(a3)} + c_W C_{uW}^{(a3)} \right\},
    \label{eq:dipole-Z-RL}
    \end{align}
where $a = 2$ denotes the charm quark. Note that all of these combinations can be complex. $\mathcal{C}_{\varphi q}^{-(3+a)}$ and  $\mathcal{C}_{\varphi u}^{(3+a)}$ lead to production of left-handed and right-handed top quarks, respectively, through a vector-like coupling to $Z$. On the other hand, $\mathcal{C}_{uZ}^{(3a)}$ and $\mathcal{C}_{uZ}^{(a3)}$ lead to production of left-handed and right-handed top quarks, respectively, through a dipole-like coupling to $Z$. In the following sections, we will present results for each of them independently.

In addition, the following {4-fermion} operators can contribute to our signal eq.\eqref{eq:single-top} without involving a propagating $Z$ boson:
\begin{eqnarray}
    \label{eq:4-fermionFCNC}
    \mathcal{O}_{lq}^{1(ijkl)} = \left( \bar{l}_i \gamma_{\mu} l_j \right) \left( \bar{q}_k \gamma^{\mu} q_l \right), \\
    \mathcal{O}_{lq}^{3(ijkl)} = \left( \bar{l}_i \gamma_{\mu} \tau^I l_j \right) \left( \bar{q}_k \gamma^{\mu} \tau^I q_l \right), \\
    \mathcal{O}_{lu}^{(ijkl)} = \left( \bar{l}_i \gamma_{\mu} l_j \right) \left( \bar{u}_k \gamma^{\mu} u_l \right), \\
    \mathcal{O}_{eq}^{(ijkl)} = \left( \bar{e}_i \gamma_{\mu} e_j \right) \left( \bar{q}_k \gamma^{\mu} q_l \right), \\
    \mathcal{O}_{eu}^{(ijkl)} = \left( \bar{e}_i \gamma_{\mu} e_j \right) \left( \bar{u}_k \gamma^{\mu} u_l \right), \\
    \mathcal{O}_{lequ}^{1(ijkl)} = \left( \bar{l}_i e_j \right) \varepsilon \left( \bar{q}_k u_l \right), \\
    \mathcal{O}_{lequ}^{3(ijkl)} = \left( \bar{l}_i \sigma_{\mu \nu} e_j \right) \varepsilon \left( \bar{q}_k \sigma^{\mu \nu} u_l \right). \label{eq:4-fermionFCNC-end}
\end{eqnarray}
{Here $\epsilon$ is the antisymmetric Levi-Civita tensor.}
{We pick $kl=(32)$ as to pick $tc$ couplings and $i=j=e$ or $i=j=\mu$ for $e^+e^-$ and $\mu^+\mu^-$ colliders, respectively. More precisely,} we pick the following linear combination of Wilson coefficient of the operators above to study 
\begin{gather}
    \mathcal{C}_{lq}^{-(1,3+a)} \equiv \left( C_{lq}^{1(113a)} - C_{lq}^{3(113a)} \right), \label{eq:4-fermion-operator}
\end{gather}
where $a=2$ denotes the charm quark. {Changes in the chirality and the $SU(2)$  structure of the operators may result in slightly different result, e.g. due to the background rates not being the same for the different chiralities. We do not explore these details in the present work, as they should lead to no change in our conclusion about the significance of testing these operators with colliders. On the contrary, the bounds these operators may receive from other observations, e.g. in flavor~\cite{Fox:2007in}, could   vary significantly depending on the details of the 4-fermion interaction under study.}

\section{Current LHC bounds: recast HVT model searches for Randall-Sundrum \label{sec:recast}}

In the Randall--Sundrum (RS) model, all Standard Model (SM) particles arise as zero modes of five-dimensional (5D) fields. Consequently, each SM particle possesses a tower of heavier Kaluza--Klein (KK) excitations, with masses typically set by a common scale, denoted \( M_{KK} \).

In particular, the KK excitations of the \( Z \) boson, denoted \( Z_{KK} \), have flavor-violating couplings to SM fermions in the mass eigenbasis. After electroweak symmetry breaking (EWSB), the SM \( Z \) boson mixes with \( Z_{KK} \), leading to tree-level flavor-changing neutral current interactions involving the top quark. {The RS model  produces fermion current-fermion current and Higgs current-fermion current operators from tree-level exchange of KK modes or mixing between KK modes and SM modes as main effects in the effective 4D theory. Dipole operators are usually suppressed because they arise at loop level in the underlying 5D Lagrangian. %
Each type of operators can be probed in different type of experiments, thus the importance of each class of operators to put a bound on the model will be discussed for each collider presented in the following.}

One of the most studied effects so far is the a rare decay mode \( t \to Z c \) that is originated by tree-level NP exchange that generates Higgs current-fermion current operators eqs.\eqref{eq:higgs-current-fermion-current}-\eqref{eq:higgs-current-fermion-current-R}. The branching ratio for this decay, computed in~\cite{Agashe:2006wa}, depends on \( M_{KK} \) as:
\begin{equation}
    \textrm{Br}( t \to Z c) \sim 10^{-5} \left( \frac{3 \, \textrm{TeV} }{M_{KK}}\right)^4 \left (\frac{(U_R)_{23}}{0.1} \right)^2, \label{eq:airen-BRt2Zc-RS}
\end{equation}
where \( (U_R)_{23} \) encodes the degree of flavor violation in the \( Z_{KK} \) couplings to right-handed up-type quarks. In the minimal RS model without additional flavor symmetries, \( (U_R)_{23} \) is fixed to an \( \mathcal{O}(1) \) quantity~\cite{Agashe:2004cp}:
\begin{equation}
    (U_R)_{23} \sim \frac{m_c}{m_t \lambda_{CKM}^2} \approx 0.1,
\end{equation}
with \( m_c \) and \( m_t \) being the charm and top quark masses, and \( \lambda_{CKM} \) the Wolfenstein parameter of the CKM matrix. Therefore, the prediction for \( \textrm{Br}( t \to Z c) \) becomes mainly a function   of \( M_{KK} \), or equivalently, the mass of the KK \( Z \) boson, \( M_{Z_{KK}} \).
{
It is worth noting that scenarios with exotic flavor symmetry structures could allow for more flavor violation in the top sector while evading strong bounds from flavor non-conserving and flavor conserving observables. In such scenarios, $(U_R)_{23}$ could be as large as $\sim \mathcal{O}(1)$. For a review in the context of composite Higgs, which can be widely applicable to RS models, see~\cite{Glioti:2024hye}.}
 
The mixing between the SM gauge bosons and their KK counterparts also leads to deviations in electroweak precision observables. Consequently, precision measurements from LEP impose stringent constraints on the KK mass scale. In the 2013 Snowmass report~\cite{TopQuarkWorkingGroup:2013hxj}, the branching ratio \( \textrm{Br}( t \to Z c) \) was evaluated assuming \( M_{KK} = 3\, \textrm{TeV} \), which was the lowest value consistent with LEP data at that time.

Since then, the LHC has performed extensive searches for heavy vector resonances in models that   resemble the RS framework. In this section, we recast the null results of those searches to derive updated bounds on \( M_{KK} \), and subsequently provide an improved upper limit on \( \textrm{Br}( t \to Z c) \) consistent with current experimental data.

In the minimal RS setup, all first-level KK modes are approximately degenerate, i.e., they appear at a common scale \( M_{KK} \). This implies that constraints on the KK states responsible for FCNCs (e.g., \( Z_{KK} \)) can be inferred from bounds on other KK modes, such as the KK gluon \( g_{KK} \), which is typically more accessible at the LHC. 

However, it is possible to decouple the scale at which first KK mode appear for various fields by adding brane localized kinetic terms~\cite{Davoudiasl:2002ua}. This flexibility allows one to decouple the most visible KK resonances, such as the KK gluon, from the \( Z_{KK} \) responsible for FCNCs, by pushing their masses higher by an $\mathcal{O}(1)$ factor. On the other hand, the KK modes of the \( W \) and \( Z \) bosons are more closely related, as they belong to the same SU(2) triplet. Hence, bounds on \( Z_{KK} \) also apply to \( W_{KK} \), and vice versa.

For this reason, we focus on constraints from CMS and ATLAS searches targeting both \( W_{KK} \) and \( Z_{KK} \). In the RS model, these resonances predominantly decay into $t$, $H$ and longitudnal $W$ and $Z$, giving $t\bar{t}$, $t\bar{b}/\bar{t}b$, (\( WW, ZZ \))$=VV$ or (\( WH, ZH \))=$VH$ final states. 
Given the limited sensitivity of the \( t\bar{t} \) and \( t\bar{b} \) final states that we found, the description of our analysis focuses exclusively on the more constraining \( VV \) and \( VH \) channels.

The branching ratios into specific final states depend on \( M_{KK} \) and additional model-dependent parameters, as discussed in~\cite{Agashe:2008jb, Agashe:2007ki}. Less minimal RS constructions can in principle change the branching ratios by factors of order 2, e.g. by triggering new decay of RS vectors into other RS states. That is a highly model dependent variation on the branching ratios of interest. For the sake of simplicity, we consider a representative range of branching ratios for each mode, inspired by ~\cite{Agashe:2008jb, Agashe:2007ki}, as summarized in \Cref{tab:branching_ratio_RS}.
\begin{table}[t!]
\centering
\begin{tabular}{|c|c|c|}
  \hline
  \textbf{Resonance} & \textbf{Decay Channel} & \textbf{Branching Fraction} \\ \hline
  \multirow{3}{*}{$W_{KK}$} & $WH$ & $20\%-50\%$ \\ \cline{2-3}
                    & $WZ$ & $0\%-20\%$ \\ \cline{2-3}
                    & $t\bar{b}/\bar{t}b$ & $0\%-80\%$ \\ \hline
  \multirow{2}{*}{$Z_{KK}$} & $ZH$ & $20\%-80\%$ \\ \cline{2-3}
                    & $WW$ & $<10\%$ \\ \hline
  \multirow{2}{*}{$A_{KK}$} & $t\bar{t}$ & $50\%-70\%$ \\ \cline{2-3}
                    & $WW$ & $20\%-50\%$\\ \hline
\end{tabular}
\caption{Dominant decay modes of various EW gauge KK modes and corresponding range of branching fractions. For completeness, we show the branching ratios for decays involving top quark, even though the limits from \( t\bar{t} \) and \( t\bar{b} \) are weaker compared to those from $VV/VH$ final state.}
\label{tab:branching_ratio_RS}
\end{table}

CMS and ATLAS have conducted direct searches for a triplet of vector bosons \cite{CMS:2022pjv, ATLAS:2020fry, ATLAS:2022enb} decaying into 
VV/VH final states and have found no evidence of new physics. However, the reported limits do not directly apply to Randall–Sundrum (RS) models, as the analyses are based on a simplified framework known as the Heavy Vector Triplet (HVT) model~\cite{Pappadopulo:2014qza}. A variant of this model, referred to as HVT-B,   is close to Kaluza–Klein (KK) electroweak resonances in RS scenarios, though there are a few key differences.

In contrast to the RS model, the HVT-B branching fractions for decay into $VV$ and $VH$ are roughly $50\%$.
Additionally, the couplings of the new vectors to SM fermions is quite different across the two models leading to slightly different production cross-sections at LHC. {Most importantly, in RS both left- and right-handed quarks couple to the heavy vectors, and the intensity of these couplings differ between $u$ and $d$ quarks, while for HVT only left-handed quarks couple to the new vectors with equal coupling strength.} Taking into account these differences, we recast the results from CMS and ATLAS by appropriately reweighting each channel by the appropriate parton luminosity. We use results from analysis of full Run2 data, corresponding to about ${138}{\textrm{fb}^{-1}}$ luminosity. The results are compiled in \Cref{fig:cms_results_rs} and \Cref{fig:cms_results_rs-2} in Appendix~\ref{app:bounds-all}. 

\begin{figure}[htbp]
  \centering
  
  
  \begin{subfigure}[b]{0.7\textwidth}
    \includegraphics[width=\textwidth]{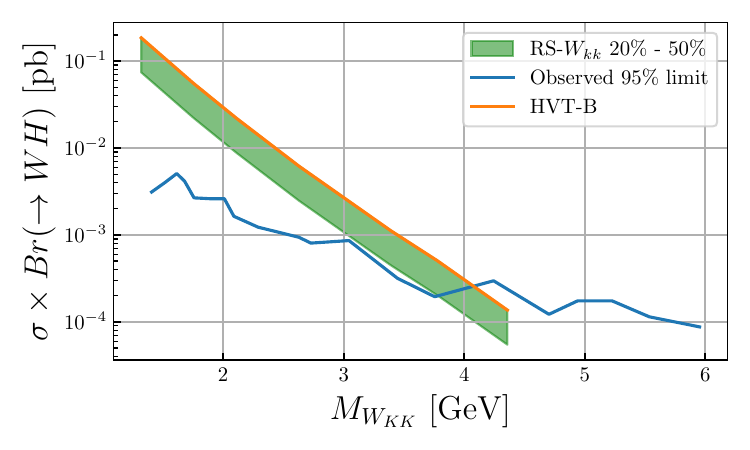}
    \label{fig:sub3}
  \end{subfigure}

  \caption{$95\%$ confidence level (C.L.) limits (blue) on the production of $W_{KK}$ decaying    into $WH$ final states as a function of mass, based on the results of~\cite{CMS:2022pjv}. 
    The orange lines represent the predictions from the HVT-B benchmark model~\cite{Pappadopulo:2014qza} with $g_V = 3$. 
    The green bands indicate the expected cross sections for vector resonances production at the 13~TeV LHC in the RS model, computed using the branching fractions listed in \Cref{tab:branching_ratio_RS}.}
  \label{fig:cms_results_rs-single}
\end{figure}

We find that the strongest bound on EW gauge KK resonances come from $WH$ resonance search at CMS, which is $\sim3.7 \, \textrm{TeV}$ and is shown in \Cref{fig:cms_results_rs-single}. 
A naive extrapolation of this single result based on the improvement with luminosity obtained by ATLAS and CMS in analyses using the full dataset compared to partial Run2 data set analysis leads us to expect this bound to improve to $4.5\, \textrm{TeV}$ for a single experiment with the luminosity $3 \textrm{ab}^{-1}$ of the HL-LHC, thus putting the combined limit from two experiments well above $5\, \textrm{TeV}$ (assuming no correlation between them). {Rescaling the $138 \textrm{fb}^{-1}$ results by PDF luminosity using  \href{http://collider-reach.web.cern.ch/}{ColliderReach}\footnote{\href{http://collider-reach.web.cern.ch/}{http://collider-reach.web.cern.ch/}} gives slightly more stringent bounds at $3 \textrm{ab}^{-1}$, but we do not use them in the following as to remain  conservative about the reach of the HL-LHC.}

\bigskip

The present limits reach higher masses for the vectors than what was probed at LEP. Further updates on direct searches for the vector bosons responsible for FCNC $Z$ boson couplings at the HL-LHC will push the prediction for $BR(t\to c Z)$ at levels around $1.2\cdot 10^{-6}$, according to eq.~\eqref{eq:airen-BRt2Zc-RS} for $M_{KK}>5{\rm ~TeV}$.

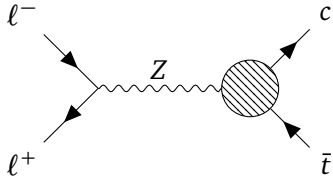
\begin{figure}
    \centering
\begin{tikzpicture}
  \begin{feynman}
    \vertex (mu) at (-1, 1) {$\ell^-$};
    \vertex (mubar) at (-1, -1) {$\ell^+$};
    \vertex (i) at (0, 0);
 
    \vertex (b) at (3,1) {${c}$};
    \vertex (bbar) at (3,-1) {$\bar t$};
    \vertex[blob] (f) at (2,0) {};

    \diagram*{
      (mu) -- [fermion] (i),
      (i) -- [fermion] (mubar),
       (i) -- [photon, edge label=\( Z\)] (f),
       (f) -- [fermion] (b),
       (bbar) -- [fermion] (f),
    };


  \end{feynman}
\end{tikzpicture}
    \caption{BSM contribution to lepton annihilation into a $tc$ pair from a $Z$ boson FCNC coupling.}
    \label{fig:ll2tc}
\end{figure}

This motivates the pursuit of processes involving FCNC $Z$ boson couplings that can probe $BR(t\to c Z)<10^{-6}$. A particularly effective probe is the production of $tc$ pairs from scattering mediated by $Z$ bosons, e.g. in Drell-Yan productions or in lepton annihilations shown in Fig.~\ref{fig:ll2tc}. 

\section{$\ell^+ \ell^- \to t c$ \label{sec:singletop}}
In this section we explore the sensitivity of $\ell^+ \ell^-$ colliders to two fermion flavor changing coupling $Ztc$ via single top production $\ell^+ \ell^- \to t c$. Following Ref.~\cite{Shi:2019epw} we discuss operators eq.~(\ref{eq:higgs-current-fermion-current})-(\ref{eq:dipole-B}), that   involve a propagating $Z$ boson, plus the four-fermion contact interactions which can lead to the same signal from Eqs.~\eqref{eq:4-fermionFCNC}-\eqref{eq:4-fermionFCNC-end}.

In the following we pursue a simplified approach presenting some purely cut-based analyses. These are accompanied by BDT results that are useful to assess the margin of improvement over our simplified analysis. Detector effects are included in a simplified manner by applying suitable amounts of noise to the simulated datasets, leaving for the future further studies of the detailed effects from detector reconstruction.

\subsection{Cut based analysis for the HTE factory \label{sec:htefactory}}
The dimension-6 operators in Eqs.~(\ref{eq:higgs-current-fermion-current})-(\ref{eq:dipole-B}) and Eqs.~(\ref{eq:4-fermionFCNC})-\eqref{eq:4-fermionFCNC-end} give rise to the following signal
\begin{equation}
 \ell^+ \ell^- \to \bar c t \to \bar c b \ell^+ \nu    \label{eq:signal-final-state},
\end{equation}
giving two jets, of which one is a $b$-jet, and a lepton in the final state. This process in the Standard Model can be mimicked by
\begin{equation}
 \ell^+ \ell^- \to W^+ W^- \to \bar c s \ell^+ \nu    \label{eq:background-final-state},
\end{equation}
when one of the light jets is mis-tagged as a $b$-jet. We assume a $b$-tagging efficiency of 80\% for true $b$-jets, while $c$-jets are misidentified as $b$-jets with a probability of 10\%:
\begin{equation}
    \epsilon_{b\to b}=0.8\quad {\rm and} \quad \epsilon_{c\to b}=0.1 \,. \label{eq:efficiency-fakes}
\end{equation}
There is also a chance that the jet from the $s$ quark is mislabeled as a $b$-jet but it is very small, hence we neglect it. The rate of $WW$ background in \cref{eq:background-final-state} is 0.09~pb after the $c\to b$ fake probability of \cref{eq:efficiency-fakes}.

{In addition to the semi-leptonic signal \cref{eq:signal-final-state} we could consider a fully hadronic signal. To the best of our knowledge this has not been considered in the earlier literature on $e^+e^-$ FCNC studies, thus we do not consider it in the present work. We remark in any case that the clean collision environment of $e^+e^-$ machines may enable a fruitful study of this final state. In the context of high energy machines that we discuss in the following this might be even more relevant, as the hadronic top quark decay products would make a characteristic ``boosted top jet'', whose rate can be added to the semi-leptonic process we study, and whose features can be used to gain more background rejection.}

We generate all parton-level processes at LO using \texttt{MG5\_aMC@NLO}\cite{Alwall:2014hca}. We use $\texttt{dim6top}$ UFO model\cite{Aguilar-Saavedra:2018ksv}\footnote{ Available at \href{https://feynrules.irmp.ucl.ac.be/wiki/dim6top}{https://feynrules.irmp.ucl.ac.be/wiki/dim6top}.} to generate signal events by turning on appropriate flavor-changing couplings. Crucial detector effects are implemented as needed and will be detailed later. 

We impose the following pre-selection criteria: jets are required to satisfy $p_T > 20$ GeV and $|\eta| < 2.5$, while leptons must fulfill $p_T > 10$ GeV and $|\eta| < 2.5$ that are expected to be a good approximation of the acceptance of future collider detectors (see for instance Refs.~\cite{IDEAStudyGroup:2025gbt,MAIA:2025hzm} for recent updates on possible detector concepts). Additionally, events must exhibit missing transverse momentum, $p_T^{\textrm{miss}} > 30$ GeV. Events passing these pre-selection cuts are further refined using two kinematic variables, $m_{t, \textrm{recoil}}$ and $m_{jj}$, defined as follows. The quantity $m_{t, \textrm{recoil}}$ represents the reconstructed mass of the top quark in signal events, determined using the charm quark energy (energy of the non-b-tagged jet) ($E_c$) via the  relation:
\begin{equation} m_{t, \textrm{recoil}}^2 = |(\sqrt{s}, \vec{0}) - p_c|^2 = s - 2\sqrt{s}E_c\, .\end{equation}

\begin{figure}[t!]
    \centering
    \includegraphics[width= \linewidth]{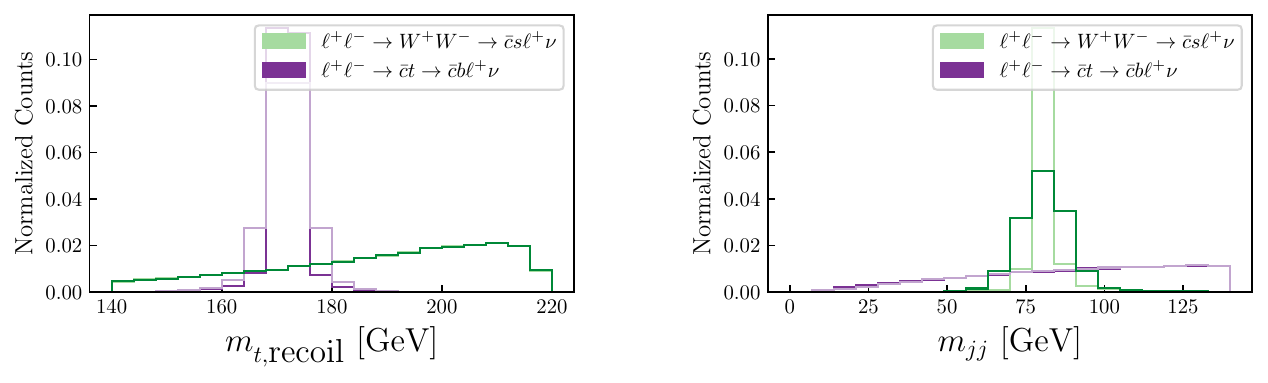}
    \caption{Distribution of $m_{t, \textrm{recoil}}$ (left) and $m_{jj}$ (right) of signal and background events at the HTE factory. The darker and lighter shades correspond to 1\% and 5\% gaussian energy smearing for the jets and the leptons.
    \label{fig: hte_mtmw}}
\end{figure}

The variable $m_{jj}$ corresponds to the invariant mass of the two-jet system in the event. In the background this is expected to resonate around $m_W$, up to detector effects.
The two variables are sensitive to the precision with which the energy of the jets are measured in the detector. Therefore, we implement energy smearing for the jets. \Cref{fig: hte_mtmw} shows the spectrum of the two variables for signal and background events for different levels of gaussian smearing 1\% and 5\%. As expected, $m_{t, \textrm{recoil}}$ peaks at $m_t = 173 \gev$ for signal events and $m_{jj}$ peaks at $80~\gev$ for background events. Fig.~\ref{fig: hte_mtmw} motivates the following final set of cuts, also adopted in Ref.~\cite{Shi:2019epw}:
\begin{equation}
    m_{t, \textrm{recoil}} < 180 \gev, \quad m_{jj}>100 \gev \,\, \textrm{ or } m_{jj}< 80 \gev \,. 
    \label{eq:cuts}
\end{equation}
We stress that Ref.~\cite{Shi:2019epw} used extra selections, e.g. on $E_c$, with thresholds that are tailored to the $e^+e^-$ 240~GeV center-of-mass case. As we will be interested in other center-of-mass energies as well, we adopt as our baseline selection \Cref{eq:cuts}, and will consider more refined selections through at BDT that can provide an automated optimal selection for each center-of-mass energy.

\begingroup
\renewcommand{\arraystretch}{1.5}
\newcommand{\tablesize}{\Large}
\begin{table}
\centering

\resizebox{\textwidth}{!}{%
\begin{tabular}{|l|ccc|ccc|ccc|ccc|}
\hline
\multirow{3}{*}{Coeff.} 
  & \multicolumn{6}{c|}{HTE factory} 
  & \multicolumn{6}{c|}{10 TeV $\mu$-coll.} \\
\cline{2-13}
  & \multicolumn{3}{c|}{Cut-based} 
  & \multicolumn{3}{c|}{BDT} 
  & \multicolumn{3}{c|}{Cut-based} 
  & \multicolumn{3}{c|}{BDT} \\
\cline{2-13}
  &   [$\text{TeV}^{-2}$] & BR $\times 10^{6}$ & $M_{KK}$~[TeV]
  &   [$\text{TeV}^{-2}$] & BR $\times 10^{6}$ & $M_{KK}$~[TeV]
  &   [$\text{TeV}^{-2}$] & BR $\times 10^{6}$ & $M_{KK}$~[TeV]
  &   [$\text{TeV}^{-2}$] & BR $\times 10^{6}$ & $M_{KK}$~[TeV] \\
\hline

$\mathcal{C}_{\varphi u}^{-(3+2)}$ & \tablesize 0.4  & \tablesize 8.5 & \tablesize 1.9 & \tablesize 0.2  & \tablesize 2.1  & \tablesize 2.7 & \tablesize 1.4 & \tablesize 100 & \tablesize 1.1 & \tablesize 1.1 & \tablesize 65 & \tablesize 1.2  \\
$\mathcal{C}_{\varphi q}^{(3+2)}$  & \tablesize 0.4  & \tablesize 8.5 & \tablesize 0.5 & \tablesize 0.2  & \tablesize 2.1  & \tablesize 0.8 & \tablesize 1.3 & \tablesize 90 & \tablesize 0.3 & \tablesize 1.1 & \tablesize 65 & \tablesize 0.3 \\
$\mathcal{C}_{uZ}^{(23)}$          & \tablesize 0.1  & \tablesize 1.7 & \tablesize 0.23 & \tablesize 0.06 & \tablesize 0.6 & \tablesize 0.29 & \tablesize 0.012 & \tablesize 0.025 & \tablesize 0.66 & \tablesize 0.01 & \tablesize 0.017 & \tablesize 0.73 \\
$\mathcal{C}_{uZ}^{(32)}$          & \tablesize 0.1  & \tablesize 1.7 & \tablesize 0.14 & \tablesize 0.06 & \tablesize 0.6 & \tablesize 0.19 & \tablesize 0.013 & \tablesize 0.029 & \tablesize 0.4 & \tablesize 0.01 & \tablesize 0.017 & \tablesize 0.46 \\

$\mathcal{C}_{lu}^{(2232)}$      & \tablesize {0.05} & \tablesize 0.005 & \tablesize \ch{1.2} & \tablesize 0.03 & \tablesize 0.002 & \tablesize \ch{1.6} & \tablesize 8$\times 10^{-5}$ & \tablesize <0.001 & \tablesize \ch{31} & \tablesize 7$\times 10^{-5}$ & \tablesize <0.001 & \tablesize \ch{33} \\

$\mathcal{C}_{eu}^{(2232)}$      & \tablesize {0.05} & \tablesize 0.005 & \tablesize \ch{1.2} & \tablesize 0.03 & \tablesize 0.002 & \tablesize \ch{1.6} & \tablesize 8$\times 10^{-5}$ & \tablesize <0.001 & \tablesize \ch{31} & \tablesize 7$\times 10^{-5}$ & \tablesize <0.001 & \tablesize \ch{33} \\

$\mathcal{C}_{lq}^{-(2,3+2)}$      & \tablesize 0.05 & \tablesize 0.005 & \tablesize 0.3 & \tablesize 0.03 & \tablesize 0.002 & \tablesize 0.4 & \tablesize 8$\times 10^{-5}$ & \tablesize <0.001 & \tablesize 7.7 & \tablesize 7$\times 10^{-5}$ & \tablesize <0.001 & \tablesize 8.2 \\

\hline
\end{tabular}%
}

\caption{ Bounds on the Wilson coefficients from our single top analysis and corresponding BSM BR at 240 GeV $e^+e^-$ collider and {10 TeV muon collider}. For dipole operators current Run-2 139/fb LHC limits are $BR(t\to c Z)< 1.3\cdot 10^{-4}$\cite{ATLAS:2023qzr,George:2023hmg}. Projected HL-LHC is expected at the level of $\text{few} \cdot 10^{-5}$, the exact number depending on assumptions on detector performances\cite{CMS-PAS-FTR-13-016, Cerri:2018ypt}. The bounds on the Wilson coefficients are also translated into limits on the $M_{KK}$
 scale (up to $\mathcal{O}(1)$ factors) in Randall–Sundrum models, following the assumptions in Refs.~\cite{Agashe:2006wa, Agashe:2004cp}. }
\label{tab:bounds-cut-based-and-BDT}
\end{table}

\endgroup

The rates of signal and background events that pass the above selection criteria are then used to derive 95\% exclusion limits for various coefficients of the operators presented in \Cref{sec:SMEFT}. We present the bounds on the various Wilson coefficients obtained using this cut-based analysis in \Cref{tab:bounds-cut-based-and-BDT}. In the table we also translate the bounds into relevant exotic top-decay branching fraction. {From this comparison we observe that search for single top production can be sensitive to couplings equivalent to branching ratios almost at the best possible sensitivity of the search of exotic decays modes, which is at the level of $O( 10^{-6})$, the exact number depending on background considerations.}

\begin{figure}[t!]
    \centering
    \includegraphics[width=0.45\linewidth]{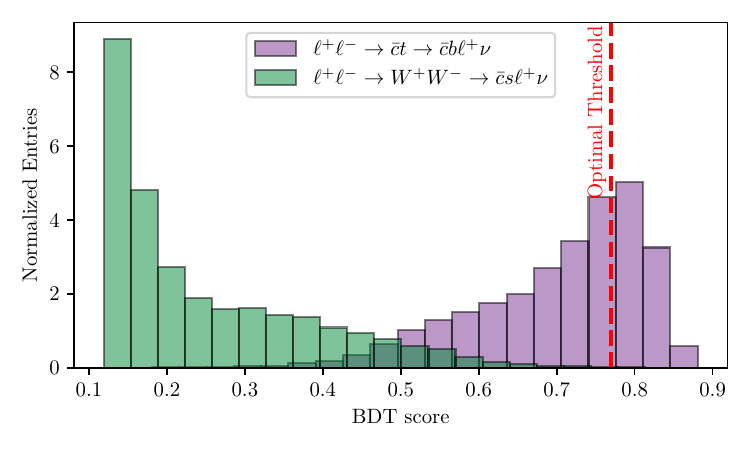}      
    \includegraphics[width=0.45\linewidth]{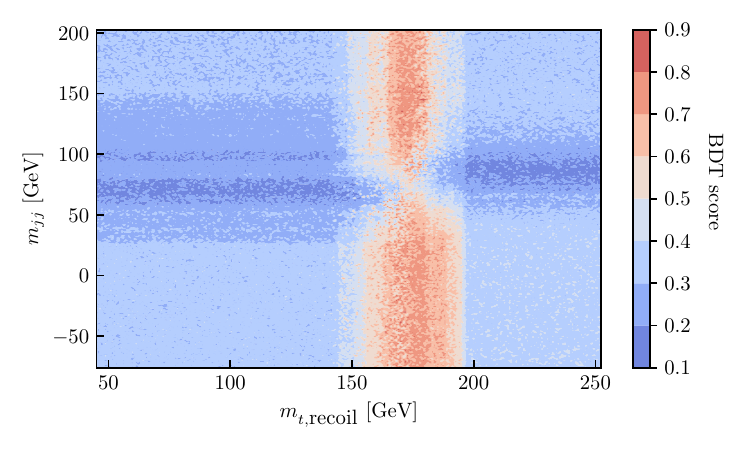}
    \caption{Left: Output of the BDT in arbitrary units. Right: BDT score in the plane $m_{t,recoil}$, $m_{jj}$.\label{fig:bdt-visualize}\label{fig:bdt-score}}
\end{figure}

Although the cuts in \cref{eq:cuts} are reasonably efficient in separating the signal and the background, they can be improved using a BDT analysis. To quantify the possible improvement we train a BDT on smeared simulated signal and background events using  as inputs the kinematic variables 
$$ m_{t, \textrm{recoil}},\; m_{jj},\textrm{ and }E_{c}\,.$$
 After training, the BDT assigns a probability score to an input event, indicating the likelihood of it being a signal event. We show the output of the BDT for the simulated data in Fig.~\ref{fig:bdt-score}.
The events above a certain threshold for the BDT output are selected. This threshold is chosen to maximize {$S/\sqrt{B}$} ratio. Additionally, the hyper-parameters of the BDT are chosen using a standard cross-validation procedure by checking for over-fitting on a slice of data unused in the training. 

The results on FCNCs obtained using this BDT analysis are shown in \Cref{tab:bounds-cut-based-and-BDT}. They are a slight improvement over the bounds from the cut-based analysis. This improvement can be attributed to the greater flexibility with which BDT can partition the input space into signal and background regions. {For illustration and to understand better how the flexibility of the BDT is reflected in the actual selection we studied the BDT score in the $ (m_{t, \textrm{recoil}}, m_{jj})$ subspace of input features.} To this end, we created a grid of points in the $ (m_{t, \textrm{recoil}}, m_{jj})$ space and randomly sampled $E_{miss}$ for each grid point from the training data distribution. These points are then used to visualize the BDT output in the right panel of \Cref{fig:bdt-visualize}. {The boundary of the red and blue regions represents the BDT's selection cuts, which vary across the 2D subspace. The better discriminative power of BDTs compared to traditional cut-based methods is testified by the non-rectangular shape of the profiles of the BDT score across the plane. The departure from a rectangular tessellation of the plane quantifies the advantage to be expected by using BDTs.}

\bigskip

\begin{figure}[t!]
    \centering
    \begin{subfigure}[b]{0.48\textwidth}
        \includegraphics[width=\textwidth]{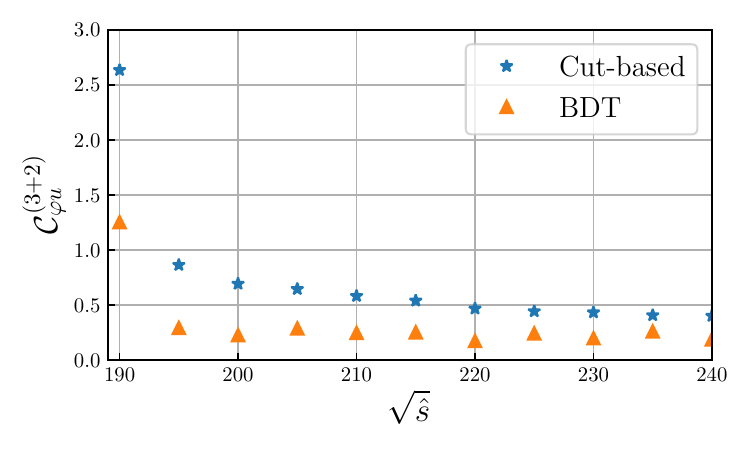}
    \end{subfigure}
    \hfill
    \begin{subfigure}[b]{0.48\textwidth}
        \includegraphics[width=\textwidth]{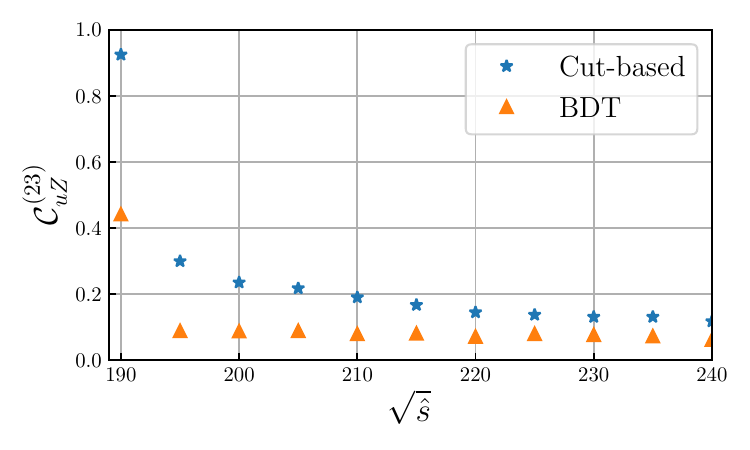}
    \end{subfigure}

    \vspace{0.5em} 

    \begin{subfigure}[b]{0.48\textwidth}
        \centering
        \includegraphics[width=\textwidth]{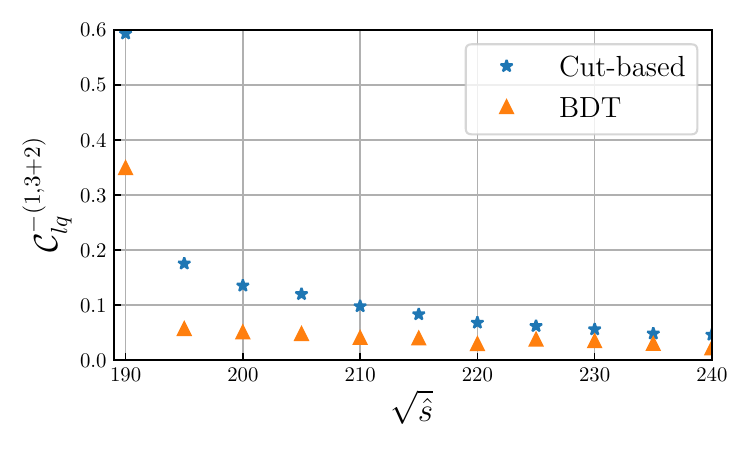}
    \end{subfigure}

    \caption{Bounds on Wilson coefficients $\mathcal{C}_{\varphi u}^{(3+2)}$, $\mathcal{C}_{uZ}^{(23)}$ and $\mathcal{C}_{lq}^{-(1,3+2)}$ as function of $\sqrt{\hat{s}}$ using a cut based analysis and BDT.
    \label{fig:varEcmBDTCuts} }
\end{figure}
The results presented so far were for fixed center-of-mass energy at 240~GeV. A simple argument based on dimensional analysis gives the following scaling of the signal cross-section with center-of-mass collider energy ($\Ecm$) for the three different operators we consider 
\begin{gather}
    \sigma_{current-current} \sim \Ecm^{-2} \,,  \label{eq:higgs-current-fermion-current-scaling}\\
    \sigma_{dipole} \sim \Ecm^0 \,,  \label{eq:dipole-scaling} \\
    \sigma_{4-fermion} \sim \Ecm^2\,, \label{eq:4-fermions}
\end{gather}
up to phase space effects. The formulae in \ref{app:dsigmadt} give the full dependence of $d\sigma/dt$ for current-current and dipole operators, which confirm this scaling, also plotted in the left panel of \Cref{fig:tc_tch_cx}. We find that background acceptance rates exhibit only a mild dependence on the center-of-mass energy, varying by less than $2\%$ between $\Ecm = 200~\gev$ and $\Ecm = 240~\gev$. {Thus, one could consider the use of different center-of-mass energies to best constrain different types of BSM contact interactions, or, as we shall do in the following \Cref{sec:muCol} the use of an altogether new higher energy machine such as a muon collider, to constrain better some interactions considered in this work.}

In the domain of circular $e^+e^-$ machine we assume that the luminosity scales roughly as $\mathcal{L}\sim \Ecm^{-4}$, which roughly tracks the projected integrated luminosity of the circular machines projects. 
From this observation we expect that  a larger total number of signal events, $N\sim \sigma\cdot \mathcal{L}$, can be observed for the current-current and dipole operators at lower collider energies than  the baseline HTE factory energy around 240 or 250~GeV, if the phase space does not suppress too much the rate. Therefore, we repeat our cut-based analysis for the HTE factory and find the new optimal BDT analysis for $\Ecm\in [200,240]~\gev$. At $\Ecm$ below the lower end of this range the phase-space closure implies $\sigma \to 0$, as $\Ecm \to m_t$, resulting in weaker bounds, as can be seen in  in \Cref{fig:varEcmBDTCuts}. From this figure we see that, using a tailored analysis as per our BDT,  comparably strong bounds can be obtained over the entire range $\Ecm\in [200,240]~\gev$.

\subsection{Analysis for  the high energy muon collider at 10~TeV \label{sec:muCol}}
\begin{figure}
  \centering
  \includegraphics[width=\linewidth]{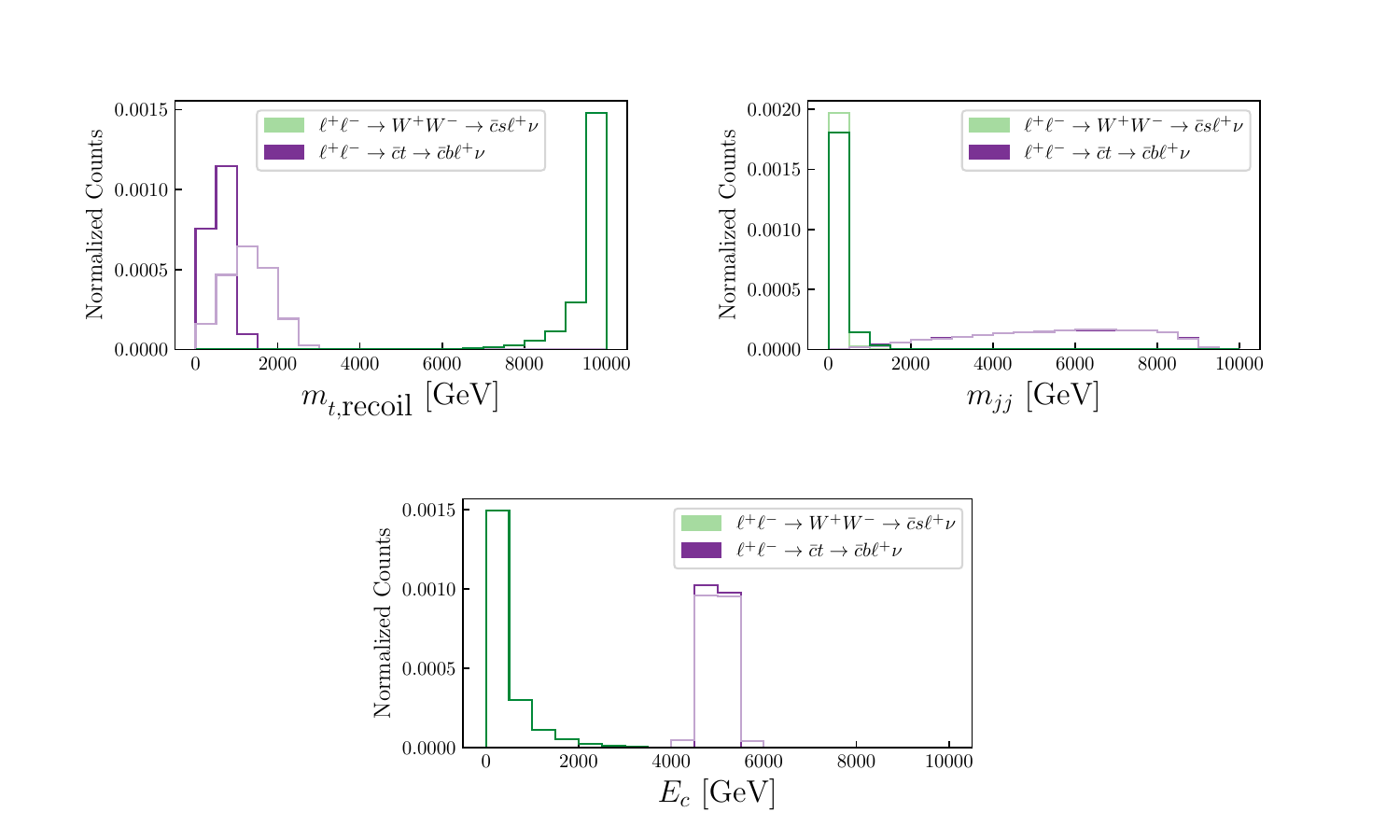}
  \caption{Distribution of $m_{t, \textrm{recoil}}$ (top-left), $m_{jj}$ (top-right) and energy of the non-b-tagged jet $E_c$ (bottom) of signal and background events at a $10\; \tev$ muon collider. The darker and lighter shades correspond to 1\% and 5\% gaussian energy smearing for the jets and the leptons.  
  \label{fig:mucSpectrum}  }
\end{figure}

The signal and backgrounds we consider are the same as for low energy HTE factory given in eq.~(\ref{eq:signal-final-state}) and (\ref{eq:background-final-state}).
The analysis follows roughly the same procedure as before, with different thresholds for the selection which are as follows:
\begin{equation}
     m_{t, \textrm{recoil}} < 4000~\gev\,, \quad m_{jj}>1000~\gev\,, \quad E_c > 2000 ~\gev\,. 
    \label{eq:10tevcuts}
\end{equation}
A few comments are in order regarding the choice of these thresholds. At a $10\, \tev$ collider, the smearing of jet energies makes it very difficult to precisely reconstruct (and discriminate between) $m_t$ and $m_W$ for signal and background events, respectively. However, as shown in \Cref{fig:mucSpectrum}, the signal and background spectra for the three kinematic variables remain clearly distinct, motivating the selection thresholds described above. Since  the signal and the background spectra are so different, our results are largely insensitive to the implementation of detector effects. {We implement  detector effects similarly to the HTE case, using 1\% and 5\% energy smearing as benchmarks. From \Cref{fig:mucSpectrum} we   see that the shapes are not very different for these two choices of resolution. In the following we quote results for 1\% energy smearing. }
The $95\%$ exclusion limits are shown in \Cref{tab:bounds-cut-based-and-BDT} assuming $10 \textrm{ ab}^{-1}$ luminosity. We also report results following a more optimized BDT analysis   as described in \Cref{sec:htefactory}.  
    
\begin{figure}
  \centering
  \begin{subfigure}[b]{0.7\textwidth}
      \includegraphics[width=\textwidth]{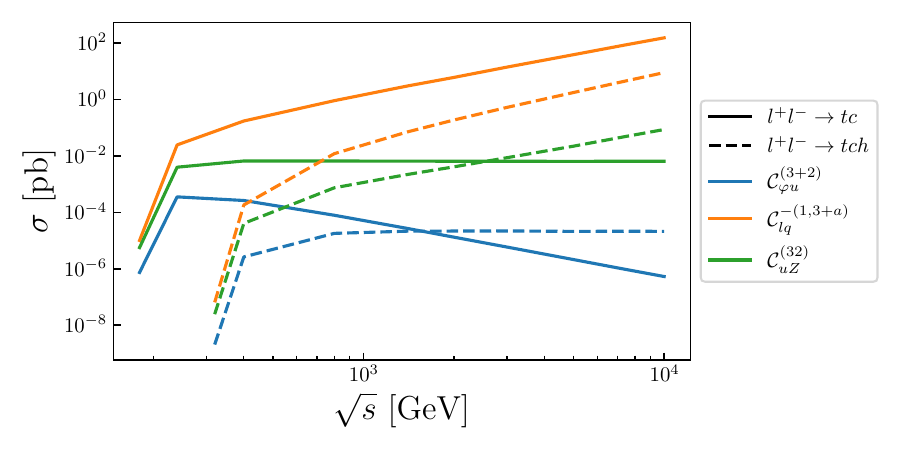}
      \label{fig: bdt_score}
  \end{subfigure}
  \caption{Cross-section for $\ell^+ \ell^- \to tc$ (solid) and $ tch $ (dashed) for the operators in the legend. Values computed with \texttt{MG5\_aMC@NLO}\cite{Alwall:2014hca}.
  \label{fig:tc_tch_cx} }
\end{figure}

As expected, a \(10\,\text{TeV}\) muon collider can significantly improve upon the current LHC limits for all considered operators, except for Higgs-current-fermion-current operators. {This is somehow expected due to the energy behavior of the cross-section mediated by these operators eq.~(\ref{eq:higgs-current-fermion-current-scaling}). } 

We expect that tighter bounds on the Higgs-current-fermion-current operators might still be achievable through processes like \(\mu^+ \mu^- \to t\bar{c}h\), whose cross-section scales more favorably with \(\sqrt{s}\). Indeed from \Cref{fig:tc_tch_cx} we can see that the total rate for $tch$ exceeds that of $tc$  for $\sqrt{s}$ above few TeV. {We leave the evaluation of the potential of this channel to future work.}

{Concerning dipole operators we find significant improvements compared to LHC limits. 
{In particular our results show that stronger bounds compared to the earlier analysis of Ref.~\cite{Ake:2023xcz} are possible. We ascribe the improvement in our results to a more refined data analysis of ours.} {Indeed, with respect to \cite{Ake:2023xcz} we   have performed a dedicated simulation of the background, instead of using rates from other results. This allows us to derive bounds in a more flexible way, working with a larger set of kinematic variables, which we use to isolate signal and background more effectively. 
Additionally, we remark that we have included some detector effects and shown that at least for the kinematic variables that we used, they do not play a significant role for the actual sensitivity of our analysis.
This verification thus reinforces our findings about the sensitivity at muon colliders.} 


{We note that  Ref.~\cite{Ake:2023xcz}   explored further and more rich final states such as $\mu^+\mu^- \to t q Z$ and $\mu^+\mu^- \to \nu \nu t q $, finding that they can contribute significant bounds at the 10~TeV muon collider, even surpassing $\mu^+\mu^-\to tq$. We leave a detailed study with detector effects for these channels and other multi-body final states such as $\mu \mu \to tch$ to future work. These additions might prove extremely helpful in a global fit exercise such as those of Refs.~\cite{Cornet-Gomez:2025jot,deBlas:2025xhe}.
}

\section{Conclusion \label{sec:conclusions}}

In this work we have considered microscopic sources of FCNC of the top quark, in particular those that give rise to $Ztc$ interactions and interactions of the four-fermion type. 
We have taken the Randall-Sundrum model as an example of model that can generate the $Ztc$  couplings at a level that could be observed at the next round of experiments, still complying with the null results of searches for the microscopic states that give rise to the FCNC interactions in that model.
To identify the relevant level of coupling strength for the $Ztc$ couplings we have derived bounds on neutral and charged  resonances in concrete RS models. To this end we have reinterpreted published LHC results on the so-called Heavy Vector Triplet model, obtaining lower limits for the RS resonances in the ballpark of 4~TeV from Run~2 data, with the prospect to obtain bounds in excess of 5~TeV after the HL-LHC. 
These results pose a target   for machines operating after the HL-LHC at the level of $Ztc$ coupling strength corresponding to $B(t\to cZ)$ around $10^{-6}$. 
{
This level of BR  can hardly be probed with a top factory producing around $10^6$ top quark pairs, thus motivated us to explore other means to probe the $Ztc$ FCNC vertex and other NP connected to flavor violation in the top quark sector\footnote{We stress that other types of NP can still show up in BSM decay modes of the top quark at a top quark factory, as recently investigated for instance in Ref.~\cite{Corcella:2025idg} with a signature agnostic method and for more specific signatures from  new physics for instance in Refs.~\cite{deBlas:2024bmz,Altmann:2025feg}}.}

Armed with this knowledge we have investigated the sensitivity of future $e^+e^-$ and $\mu^+\mu^-$ collider projects. We find that the pair production of top quark pairs may be surpassed in sensitivity by lower energy collisions in which single production of top quarks can be attained. We have investigated the center-of-mass energy range around the currently planned 240~GeV stages for $e^+e^-$ machines, finding that there is no loss of sensitivity to the $Ztc$ coupling strength if the lower energy of the collider is accompanied by an increase of the luminosity $\mathcal{L}\sim E^{-4}$.
In carrying out this exploration of runs at lower energy we  used a BDT analysis that allowed to efficiently tailor the selection to best exploit the sensitivity at each center-of-mass energy.

The analysis we have devised can be used to target other manifestations of new physics, not only a FCNC $Ztc$ coupling. We have applied it to  a set of possible SMEFT contact interactions that can give rise to single top production.  The several classes of SMEFT operators that we have analyzed are expected to give BSM effects with different behavior as one changes the center-of-mass energy of the collisions. This leads to a pattern of improvements over the projected LHC limits that can favor lower energies machines or higher energy ones depending on the specific operator. 

Operators of the type Higgs current-fermion current are probed more effectively with high intensity machines operating at lower energies, dipole-like and four-fermion operators tend to be better probed at high energy machines, with the four-fermion operators enjoying the largest benefit of the  higher energies  in enhancing the signal rates.
This reflects the  growing,  constant or increasing behavior with energy of eqs.\eqref{eq:higgs-current-fermion-current-scaling}, \eqref{eq:dipole-scaling}, and \eqref{eq:4-fermions}, for Higgs current-fermion current, dipole and four-fermion operators, respectively. {From the quantitative results of Table~\ref{tab:bounds-cut-based-and-BDT} we see that bounds on the RS KK modes mediating FCNC can be improved by a modest factor around the TeV scale at low energy machines, while high energy machines can reach up to few tens of TeV for the KK masses.} Recent studies have confirmed a reach on the SMEFT Wilson coefficients similar to our case at high energy muon colliders~\cite{Glioti:2025zpn}, thus confirming the potential for flavor physics at these machines.

It is particularly relevant  that the bounds in \cite{Glioti:2025zpn}, as well as our results, significantly exceed the reach of low energy flavor experiments, thus bringing high-energy machines at the frontier of flavor physics. {A cautionary remark is in order. In the context of RS models that we studied here we have displayed the effect of model-dependent relations between the expected naive scale of the EFT and the actual new physics particles mass scale, which can be one or two orders of magnitude smaller, due to the typical size of the flavor-violating effects. These are in our case controlled by SM parameters such as $V_{ts}$ or $m_c/m_t$ that are of order $10^{-2}$. We have also  explored the differences in the size of the effects involving left-handed and right-handed fields. The details of the RS model result in a different strength of the bounds on the underlying degrees of freedom of the UV theory obtained from each operator. We stress that such differences can be potentially be used to characterize the origin of any positive signals in these searches.}

{ Taking a wider perspective on the probe of the source of the flavor violation in the models considered here, we point out that    results from electroweak measurements at low energy machines \cite{deBlas:2944678} are expected to give the leading overall constraints on the RS KK scale. Similarly for high-energy machines  electroweak probes \cite{Buttazzo:2020uzc,InternationalMuonCollider:2025sys} will lead the overall constraint on the RS model. In view of this dominance of electroweak constraints in the study of RS phenomenology a signal in top quark flavor observables would point to a somewhat non-trivial flavor structure of the model, e.g. involving a structured sector of flavor-aware spin-1 and spin-1/2 resonances which generate small quark masses and small CKM elements from otherwise large contributions.} 

As an outlook for further studies in flavor related processes, we stress that SMEFT contact interactions can give rise to a host of different manifestations, not only the two-body processes we have studied in this work. Exploiting the patterns of the manifold manifestations of new physics is clearly a key to efficiently search for it, and, in case of observation of deviations, to understand it and characterize it. Along this path we have identified possible processes to be included for a more thorough analysis, which include $\ell^+\ell^- \to tch$ and $tc$ production from vector boson fusion. These processes have sizable cross-section at colliders exceeding few~TeV center-of-mass energy and may significantly enhance the sensitivity to operators such as the Higgs current-fermion current operators.

\section*{Acknowledgements}
We thank Gauthier Durieux and David Marzocca for valuable comments on our manuscript and Patrizia Azzi, Nuno Castro, Marina Cobal,   Mar\'ia Teresa N\`u\~nez Pardo de Vera, Rebeca Gonzalez Suarez, Fabio Maltoni, Kirill Skovpen, Marcel Vos, and Aleksander Filip Zarnecki for discussions.
RF is supported in part by the European Union - Next
Generation EU   Missione 4 - Componente 2 - Investimento 1.1 
 CUP: I53D23000950006 through the MUR PRIN2022 Grant n.202289JEW4. 


\begin{appendix}
\numberwithin{equation}{section}
\section{Current LHC Bounds - Additional Results \label{app:bounds-all}}
\begin{figure}[htbp]
  \centering
  \begin{subfigure}[b]{0.49\textwidth}
    \includegraphics[width=\textwidth]{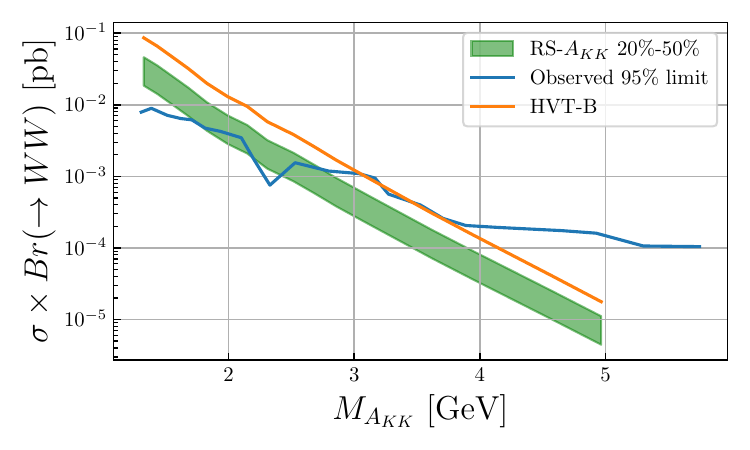}
    \label{fig:sub1}
  \end{subfigure}
  \hfill
  \begin{subfigure}[b]{0.49\textwidth}
    \includegraphics[width=\textwidth]{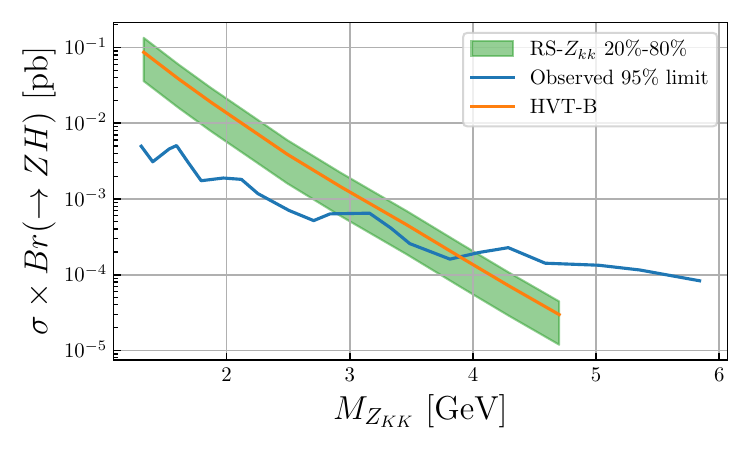}
    \label{fig:sub2}
  \end{subfigure}
  
  \vspace{0.3cm}
  
  \begin{subfigure}[b]{0.49\textwidth}
    \includegraphics[width=\textwidth]{./plots/cms_wkk_wh.pdf}
    \label{fig:sub3}
  \end{subfigure}
  \hfill
  \begin{subfigure}[b]{0.49\textwidth}
    \includegraphics[width=\textwidth]{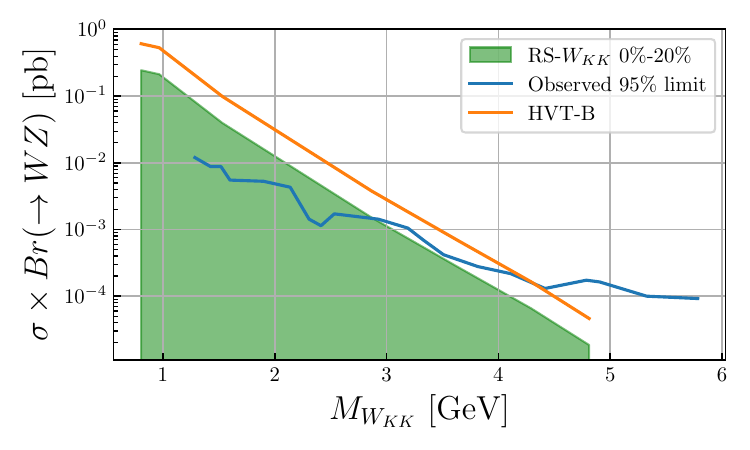}
    \label{fig:sub4}
  \end{subfigure}

  \caption{$95\%$ confidence level (C.L.) limits (blue) on the production of vector resonances decaying into $VV/VH$ final states as a function of mass, based on the CMS results of~\cite{CMS:2022pjv}. 
    The orange lines represent the predictions from the HVT-B benchmark model~\cite{Pappadopulo:2014qza} with $g_V = 3$. 
    The green bands indicate the expected cross sections for vector resonances in the RS model, computed using the branching fractions listed in \Cref{tab:branching_ratio_RS}.}
  \label{fig:cms_results_rs}
\end{figure}

\begin{figure}[htbp]
  \centering
  \begin{subfigure}[b]{0.49\textwidth}
    \includegraphics[width=\textwidth]{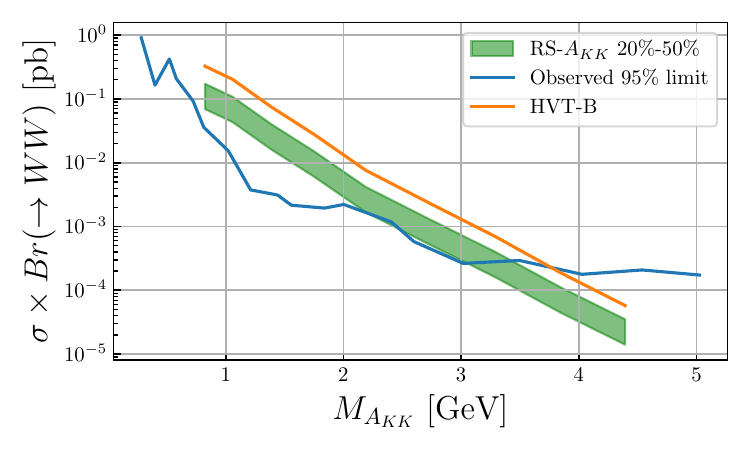}
    \label{fig:sub1}
  \end{subfigure}
  \hfill
  \begin{subfigure}[b]{0.49\textwidth}
    \includegraphics[width=\textwidth]{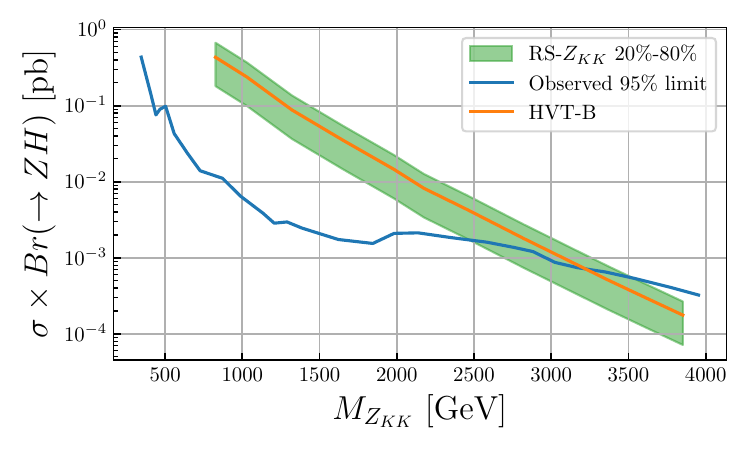}
    \label{fig:sub2}
  \end{subfigure}
  
  \vspace{0.3cm}
  
  \begin{subfigure}[b]{0.49\textwidth}
    \includegraphics[width=\textwidth]{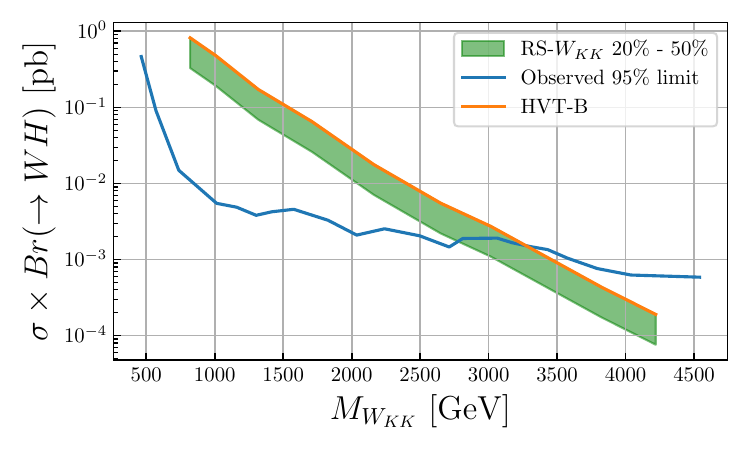}
    \label{fig:sub3}
  \end{subfigure}
  \hfill
  \begin{subfigure}[b]{0.49\textwidth}
    \includegraphics[width=\textwidth]{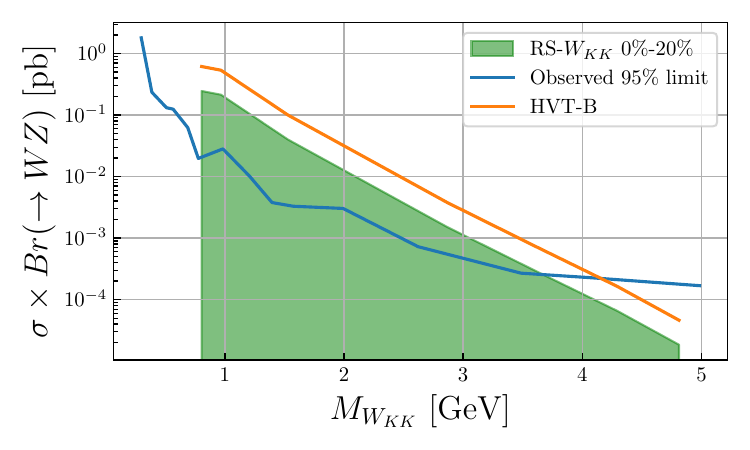}
    \label{fig:sub4}
  \end{subfigure}

  \caption{$95\%$ confidence level (C.L.) limits (blue) on the production of vector resonances decaying into $VV/VH$ final states as a function of mass, based on the ATLAS results of~\cite{ATLAS:2020fry}. 
    The orange lines represent the predictions from the HVT-B benchmark model~\cite{Pappadopulo:2014qza} with $g_V = 3$. 
    The green bands indicate the expected cross sections for vector resonances in the RS model, computed using the branching fractions listed in \Cref{tab:branching_ratio_RS}.}
  \label{fig:cms_results_rs-2}
\end{figure}

\newpage 
\section{Expressions for the signal differential cross-sections \label{app:dsigmadt}}
In the SM the flavor-changing process we are after has a negligible rate. It is zero in the approximation taken in our UFO model. Thus the signal cross-section is made of just  the contributions from amplitude squared involving SMEFT operators. The resulting  cross-section for current-current eq.~\eqref{eq:singlet-current-current} operators $d\sigma/dt$ reads:
\begin{multline}
    \frac{|\mathcal{C}_{\varphi u}^{(3+2)}|{}^2g_Z^4 v_{EW}^4}{128 \pi^2 s^2 (M_Z^2 - s)^2} \Bigg[ 4 \left( m_t^2 (t + u) - t^2 - u^2 \right) \sin^4\theta_W  \\ - 4 t (m_t^2 - t) \sin^2\theta_W + t (m_t^2 - t) \Bigg]\,,
\end{multline}
where $s,t,u$ are the Mandelstam variables, $g_Z$ is the SM $Z$ coupling, $m_t$ is the mass of the top quark, $\theta_W$ is the Weinberg angle and $v_{EW}$ is the Higgs vev. 

For dipole-like operators eq.~\eqref{eq:dipole-Z-LR} the cross-section $d\sigma/dt$ reads:
\begin{multline}
     \frac{|\mathcal{C}_{uZ}^{(32)}|{}^2g_{Z}^2 v_{EW}^2}{32 \pi^2 s (M_Z^2 - s)^2}  \Bigg[\left( m_t^2 (t + u) - 2 t u \right) \cos4\theta_W  -  \\ 4 u (m_t^2 - t) \cos2\theta_W  + m_ t^2 t + 3 m_t^2 u - 4 t u \Bigg]\,.
\end{multline}

\end{appendix}

\bibliography{SciPost_Example_BiBTeX_File.bib}


\end{document}